\documentclass[prd,nofootinbib,onecolumn,superscriptaddress]{revtex4-2}

\usepackage{amsmath, amssymb, amsthm, graphicx, epsfig, fancyhdr,epsfig, slashed, mathrsfs,bm}

\usepackage{natbib}
\usepackage{float}
\usepackage{layout}

\usepackage{mathtools} 

\usepackage{tikz,xcolor,hyperref}

\definecolor{lime}{HTML}{A6CE39}
\DeclareRobustCommand{\orcidicon}{
	\begin{tikzpicture}
	\draw[lime, fill=lime] (0,0) 
	circle [radius=0.2] 
	node[white] {{\fontfamily{qag}\selectfont \tiny ID}};
	\draw[white, fill=white] (-0.0625,0.095) 
	circle [radius=0.007];
	\end{tikzpicture}
	\hspace{-2mm}
}

\foreach \x in {A, ..., Z}{\expandafter\xdef\csname orcid\x\endcsname{\noexpand\href{https://orcid.org/\csname orcidauthor\x\endcsname}
			{\noexpand\orcidicon}}
}

\newcommand{\be}{\begin{equation}}
\newcommand{\ee}{\end{equation}}
\newcommand{\bea}{\begin{eqnarray}}
\newcommand{\eea}{\end{eqnarray}}

\begin{document}


\title{Review of strongly coupled regimes in gravity with Dyson-Schwinger approach}


\author{Marco Frasca\orcidA{}}
\email{marcofrasca@mclink.it }
\affiliation{Rome, Italy}

\author{Anish Ghoshal\orcidB{}}
\email{anish.ghoshal@sussex.ac.uk}
\affiliation{Department of Physics and Astronomy, University of Sussex, \\
Brighton,  BN1 9RH, United Kingdom}

\begin{abstract}
\textit{
We analyze various gravity theories involving de-Sitter, quadratic $\mathcal{R}^2$ and non-minimally coupled scalar in the light of application of the Dyson-Schwinger technique involving exact background solution of the Green's function. We denote specific set of solutions for the metric to move towards a quantum analysis of the theory. This kind of solutions is identified as conformally flat metric. Such a conclusion naturally arises in the use of the Dyson-Schwinger equations in the study of the Yang-Mills theory through the mapping theorem. We show a sequence of cosmological phase transitions starting from the breaking of such conformal invariance that 
can be hindered by the presence of the non-minimal coupling.
}
\end{abstract}

\maketitle


\section{Introduction}
Among the theories of gravity, Einstein's general relativity (GR) has been extraordinarily successful classical theory. It has been stood the test of time by being consistent with several observations. Its most recent success include the advent of gravitational waves (GW), which originates from coalescing black holes seen by the LiGO GW detector~\cite{LIGOScientific:2016aoc}. Along with these, we also have obtained the images of the black holes which lay at the center of our galaxy involving the M87 as seen by the Event Horizon Telescope~\cite{EventHorizonTelescope:2019dse,EventHorizonTelescope:2019ggy,EventHorizonTelescope:2022xnr,EventHorizonTelescope:2022wok,EventHorizonTelescope:2022xqj}. Not only that as we know Einstein's theory of general relativity also extends to incorporate matter fields. This forms the fundamental basis of our modern understanding of cosmology, which also includes the early universe and origin of Big Bang.

Having said these, Einstein's GR behaves pathologically and suffers non-renormalizability by known perturbative methods~\cite{Goroff:1985sz,Goroff:1985th} in the UV and be within the regime of validity of perturbation theory at length scales of the thoery much below the Planck scale in order to make sense of the quantum regime. Another aspect where GR falls short is the lack of a consistent Euclidean path integral: the Euclidean action of GR is unbounded from below, also known as the ``conformal-factor problem".  Both the non-renormalizability and the conformal-factor problem of GR could be solved by adding quadratic-in-curvature terms. Indeed, the resulting theory, which is often called quadratic gravity, is renormalizable~\cite{Weinberg:1974tw,Deser:1975nv,Stelle:1976gc,Barvinsky:2017zlx} and a simple application of the prescription~\cite{Gibbons:1994cg} to determine the Euclidean action indicates that is also free from the conformal-factor problem in a sizable region of its parameter space~\cite{Menotti:1989ms}.

In this review we analyze some gravity models that can be treated using the quantization technique of Dyson-Schwinger equations wherever the main degrees of freedom can be reduced to the solution of a scalar field theory. The idea to obtain an understanding of quantum gravity on similar concepts but different methods has been pursued in several papers as for the Lee-Wick model~\cite{Lee:1969fy,Salvio:2018kwh,Donoghue:2019fcb,Holdom:2021hlo} the corresponding fakeon prescription~\cite{Anselmi:2017ygm}, and also \cite{odins1, odins2, odins3} by the authors in string-inspired non-local ghost-free approaches to higher-derivative gravity \cite{Frasca:2020jbe,Frasca:2020ojd,Frasca:2021iip, Frasca:2022duz,Frasca:2022gdz}. An understanding for quadratic gravity was provided in Ref. \cite{Salvio:2024joi,Choudhury:2025iwg}. 

Dyson-Schwinger technique that we apply here arose from a paper by Bender, Milton and Savage~\cite{Bender:1999ek} that obtained the set of Dyson-Schwinger equations for a PT-invariant model in shape of partial differential equations. Jacobi elliptic functions turn out a powerful tool to solve such kind of equations in quantum field theories as shown in Ref.~\cite{Frasca:2015yva}. Several applications were devised in particle physics and quantum chromodynamics \cite{Frasca:2021yuu,Frasca:2021mhi,Frasca:2022lwp,%
Frasca:2022pjf,Chaichian:2018cyv,Frasca:2023qii,Frasca:2023eoj}.

We will show that, for some models of modified gravity, assuming as leading solution to the set of Dyson-Schwinger equations a conformal solution, some phase transitions at the early universe can be identified. The need for a modified model of gravity arises when one observes that for a Einstein-Hilbert action, starting with conformal solutions is not the right way to quantize them.

The paper is organized as follows: in Section 2 we introduce the Dyson–Schwinger
technique; in Section 3 we show the de Sitter solution based on that technique; in Section 4
we show this for the explicit Starobinsky theory for $R + R^2$ scenarios; In Section 5 we
quantize the scalaron field, In Section 6 we consider the effect of matter fields. In Section 7,
we discuss the non-minimal coupling case. Finally, in Section 8 we present the conclusions
and some discussion.

\medskip

\section{Dyson-Schwinger technique}

In this section we introduce a technique devised by Bender, Milton and Savage \cite{Bender:1999ek} to obtain the set of Dyson-Schwinger equations in shape of partial differential equations. The original idea arose in the study of PT-invariant quantum mechanical models. This approach has been recently extended to the more general setting of quantum field theory  \cite{Frasca:2015yva}. We consider the following partition function of a generic scalar field ($\phi$) theory in Euclidean metric
\be
Z[j]=\int[d\phi]e^{-\int d^4x[\frac{1}{2}(\partial\phi)^2+\frac{\lambda}{4}\phi^4+j\phi]},
\ee
being $j$ an arbitrary source and $\lambda$ the coupling constant entering into the self-interaction term. In order to evaluate this partition function, we need to obtain all the set of nP-correlation functions in closed form. These are given by the following functional derivatives
\be
G_n(x_1,x_2,\ldots,x_n)=\left.\frac{\delta^n\ln Z}{\delta j(x_1)\delta j(x_2)\ldots\delta j(x_n)}\right|_{j=0},
\ee
From our expression from the correlation functions the following relations follow
\bea
G_1(x_1)&=&\left.\frac{1}{Z}\frac{\delta Z}{\delta j(x_1)}\right|_{j=0}, \nonumber \\
G_2(x_1,x_2)&=&\left.\frac{\delta G_1^{(j)}(x_1)}{\delta j(x_2)}\right|_{j=0}, \nonumber \\
G_3(x_1,x_2,x_3)&=&\left.\frac{\delta G_2^{(j)}(x_1,x_2)}{\delta j(x_3)}\right|_{j=0}, \nonumber \\
&\vdots&.
\eea
The procedure evaluates the PDEs for the following
\be
G_n^{(j)}(x_1,x_2,\ldots,x_n)=\frac{\delta^n\ln Z}{\delta j(x_1)\delta j(x_2)\ldots\delta j(x_n)},
\ee
that are the correlation functions dependent on the source $j$. The starting point is given by the equation of motion
\be
-\partial^2\phi+\lambda\phi^3=j.
\ee
We want o obtain $G_1$, thus, we take the average of this equation using $Z$ yielding
\be
\label{eq:G1j}
-\partial^2G_1^{(j)}(x)+\lambda Z^{-1}\langle\phi^3\rangle=j.
\ee
We recognize here the average of the cube of the field. This can be evaluated using its definition
\be
Z G_1^{(j)}(x)=\langle\phi\rangle.
\ee
Through iteration of the derivative with respect to $j(x)$ we get
\bea
&&Z G_2^{(j)}(x,x)+Z[G_1^{(j)}(x)]^2=\langle\phi^2\rangle, \nonumber \\
&&3Z G_1^{(j)}(x)G_2^{(j)}(x,x)+Z G_3^{(j)}(x,x,x)+Z[G_1^{(j)}(x)]^3=\langle\phi^3\rangle.
\eea
These can be straightforwardly substituted in eq.(\ref{eq:G1j}) providing the first Dyson-Schwinger equation 
\be
\label{eq:G1j1}
-\partial^2G_1^{(j)}(x)+\lambda \left\{3G_2^{(j)}(x,x)G_1^{(j)}(x)+G_3^{(j)}(x,x,x)+[G_1^{(j)}(x)]^3\right\}=j.
\ee
Finally, we can take $j=0$ so that
\be
\label{eq:G1j2}
-\partial^2G_1(x)+\lambda [3G_1(x)G_2(x,x)+G_3(x,x,x)+G_1^3(x)]=0.
\ee
For $G_2$, we derive eq.(\ref{eq:G1j1}) with respect to $j(x_2)$ obtaining
\bea
&&-\partial^2G_2^{(j)}(x,x_2)+\lambda\left\{3G_2^{(j)}(x,x)G_2^{(j)}(x,x_2)+3G_2^{(j)}(x,x)G_3^{(j)}
(x,x,x_2)+\right. \nonumber \\
&&\left.G_4^{(j)}(x,x,x,x_2)+3[G_1^{(j)}(x)]^2G_2^{(j)}(x,x_2)\right\}=\delta^4(x-x_2),
\eea
and take $j=0$ again . Such a procedure can be iterated at any desired order providing the equations for higher-order correlation functions. We see that such higher-order correlation functions appear in the lower-order equations. This is a characteristic of the Dyson-Schwinger set of equations but we can provide an exact Gaussian solution for the partition function by assuming $G_{n>2}(x_1,x_1,\ldots)=0$ that is, when at least a couple of variables are identical. This choice appears consistent {\sl a posteriori} and provides the full solution to the set of Dyson-Schwinger equations just through $G_1$ and $G_2$.

In order to understand the physical meaning of this approach, we point out that the solution of the set of Dyson-Schwinger equations grants a correspondingly exact solution for the quantum field theory they are derived from. The reason for this relies on the fact that the observables of a statistical theory are completely computable when all the correlation functions are exactly known. Specifically, for a quantum field theory, this is the statement of the Lehmann–Symanzik–Zimmermann (LSZ) reduction formula \cite{Lehmann:1954rq}.

\medskip

\section{de Sitter solution}

From the preceding discussion of the Dyson-Schwinger technique applied to the non-Abelian gauge theories, it appears rather clear that some kind of conformal solution is needed (the mapping theorem in our case) to grant a direct application of the method. In the Einstein theory, we are in a very similar situation and the simplest mapping solutions one can think of are the conformally flat solutions. We know that Einstein equations are not conformally invariant and, we a standard textbook case, we show how the Dyson-Schwinger approach needs an extended gravity sector to be consistently applied. So, we consider the vacuum Einstein equations with cosmological constant $\Lambda$
\begin{equation}
\mathcal{R}_{\mu\nu} = \Lambda g_{\mu\nu},
\end{equation}
the metric having signature  $(1,-1,-1,-1)$, and restrict our attention to the conformally flat sector
\begin{equation}
g_{\mu\nu} = e^{2\phi(x)}\eta_{\mu\nu},
\end{equation}
where $\eta_{\mu\nu}$ is the Minkowski metric and $\phi(x)$ is a scalar conformal factor.
The Ricci tensor takes the form
\begin{equation}
\mathcal{R}_{\mu\nu}
= -2\partial_\mu\partial_\nu\phi
-\eta_{\mu\nu}\Box\phi
+ 2\partial_\mu\phi\partial_\nu\phi
-2\eta_{\mu\nu}(\partial\phi)^2,
\end{equation}
where $\Box = \eta^{\mu\nu}\partial_\mu\partial_\nu$ and $(\partial\phi)^2 =\eta^{\mu\nu}\partial_\mu\phi,\partial_\nu\phi$. Similarly, the Ricci scalar takes the form
\begin{equation}
\mathcal{R} = -6e^{-2\phi}\big(\Box\phi + (\partial\phi)^2\big).
\end{equation}
Inserting the Ricci tensor into $R_{\mu\nu}=\Lambda g_{\mu\nu}$ yields the tensorial equation
\begin{equation}
\label{eq:Ein}
-2\partial_\mu\partial_\nu\phi
+2\partial_\mu\phi\partial_\nu\phi
= \eta_{\mu\nu}
\big( \Box\phi + 2(\partial\phi)^2 + \Lambda e^{2\phi} \big),
\end{equation}
and taking the trace this gives the single independent equation of motion
\begin{equation}
\Box\phi + (\partial\phi)^2 + \frac{2}{3}\Lambda e^{2\phi} = 0.
\end{equation}
Using this equation into eq.(\ref{eq:Ein}), we get
\begin{equation}
\partial_\mu\partial_\nu\phi
-\partial_\mu\phi\partial_\nu\phi
  = \frac14\eta_{\mu\nu}
  \big( \Box\phi - (\partial\phi)^2 \big).
\end{equation}
This equation does not introduce new dynamics; it is a compatibility condition ensuring that the conformally flat ansatz is consistent with an Einstein metric.
Using
\begin{equation}
\partial_\mu\partial_\nu\phi - \partial_\mu\phi\partial_\nu\phi
= - e^{\phi}\partial_\mu\partial_\nu(e^{-\phi}),
\end{equation}
the compatibility condition can be written compactly as
\begin{equation}
\big( \partial_\mu\partial_\nu e^{-\phi} \big)^{\mathrm{TL}} = 0,
\end{equation}
where ``TL'' denotes the traceless part with respect to $\eta_{\mu\nu}$. Equivalently,
\begin{equation}
\partial_\mu\partial_\nu e^{-\phi}
= \frac14\eta_{\mu\nu}\Box e^{-\phi}.
\end{equation}
This implies that $e^{-\phi}$ is at most quadratic in the Minkowski coordinates and therefore generates precisely Minkowski, de Sitter, or anti--de Sitter spacetimes in conformally flat coordinates.
In order to solve the equations of motion, we select the generalized harmonic gauge granted by the condition \cite{Pretorius:2006tp} 
\begin{equation}
W^\mu[g] \equiv g^{\alpha\beta}\Gamma^\mu_{\alpha\beta}.
\end{equation}
This choice grants consistency between mathematical formulation and physics (a simple harmonic gauge would not be enough). Then, for the conformal metric one finds the identity
\begin{equation}
W^\mu[g] = -2e^{-2\phi}\partial^\mu\phi.
\end{equation}
In the conformally flat reduction, 
the gauge function is \emph{completely determined} by the scalar field $\phi$. It is not an independent equation and does not impose further constraints on the dynamics.
Then, we look for a solution depending only on the conformal time $\eta$,
\begin{equation}
\phi = \phi(\eta),
\end{equation}
yielding for The compatibility condition
\begin{equation}
(\partial_\mu\partial_\nu e^{-\phi})^{\mathrm{TL}} = 0
\end{equation}
then implies
\begin{equation}
\partial_\eta^2 e^{-\phi} = 0,
\end{equation}
whose general solution is
\begin{equation}
e^{-\phi(\eta)} = A\,\eta + B,
\end{equation}
with constants $A,B$. Without loss of generality, one may set $B=0$ by a shift of $\eta$, and define $A=H>0$, so that
\begin{equation}
\phi(\eta) = -\ln(H\eta).
\end{equation}
Inserting this expression into the scalar equation of motion
\begin{equation}
\Box\phi + (\partial\phi)^2 + \frac{2}{3}\Lambda e^{2\phi} = 0,
\end{equation}
one finally finds
\begin{equation}
H^2 = \frac{\Lambda}{3}.
\end{equation}
The resulting metric is therefore
\begin{equation}
 ds^2 = \frac{1}{(H\eta)^2}
 \left(-d\eta^2 + d\vec{x}^2\right),
\end{equation}
which is de Sitter spacetime written in conformally flat (Poincar\'e) coordinates.
The relevant conclusion from this textbook example is that the scalar field equation and the constraint force the solution to be a de Sitter or Minkowski one and a direct quantization using the Dyson-Schwinger method appears impossible for the Einstein equations in vacuum. We need to extend the theory adding further terms to the standard Einstein-Hilbert action.  

\medskip

\section{$\mathcal{R}+\mathcal{R} ^2$ Theory}

With the aim to enlarge the gravity sector, the simplest step is to consider the action \cite{Starobinsky:1980te,Mukhanov:1981xt,Kaneda:2010ut}
\begin{equation}
\label{eq:SS}
S = -\frac{M_p^2}{2} \int d^4x \sqrt{-g} \left(\mathcal{R} + \frac{\mathcal{R}^2}{6 M^2}\right),
\end{equation}
where $M_p$ represents the Planck scale, $\mathcal{R}$ is the Ricci scalar and M$^{-1}$ denotes the scale at which the $\mathcal{R}^2$ term comes into play. The action (\ref{eq:SS}) can be re-written in terms of the interaction between gravity and a scalar field $\chi$ after performing the transformation and so, moving to the Einstein frame, \cite{Whitt:1984pd,Kehagias:2013mya}
\be
g_{\mu\nu}\rightarrow e^{-\sqrt{\frac{2}{3}}\frac{\chi}{M_{\rm p}}}g_{\mu\nu},
\ee
where $g_{\mu\nu}$ is a generic metric. This is the analogous of or mapping theorem for Yang-Mills theory provided we consider some specific kind of metrics.
The result is the following action \cite{Salvio:2018crh}:
\be
  S =\int {\rm d}^4 x\sqrt{-g}\left[-\frac{M_{\rm p}^2}{2}\mathcal{R}+\frac{1}{2}g^{\mu\nu}\partial_\mu\chi\partial_\nu \chi+\frac{3}{4} M_{\rm p}^2 M^2\left(1-e^{-\sqrt{\frac{2}{3}}\frac{\chi}{M_{\rm p}}}\right)^2\right]. \label{R3}
\ee
The equations of motion are then
\bea
&&\mathcal{R}_{\mu\nu} = \frac{1}{M_p^2} \left\{ 
 \partial_\mu \chi \, \partial_\nu \chi
- \tfrac{1}{2} g_{\mu\nu} \, g^{\alpha\beta} \partial_\alpha \chi \, \partial_\beta \chi
- \tfrac{3}{4} M_{\rm p}^2 M^2 g_{\mu\nu} \left(1 - e^{-\sqrt{\tfrac{2}{3}} \tfrac{\chi}{M_{\rm p}}}\right)^2
\right\}
+\frac{1}{2}g_{\mu\nu}{\cal R},
\nonumber \\
&&\Box_g\chi= \sqrt{\frac{3}{2}} M^2 M_p e^{-\sqrt{\frac{2}{3}} \frac{\chi}{M_p}} \left(1 - e^{-\sqrt{\frac{2}{3}} \frac{\chi}{M_p}}\right),
\eea
where $\Box_g=-\frac{1}{\sqrt{-g}}\partial_\mu g^{\mu\nu}\sqrt{-g}\partial_\nu$ is the Laplace-Beltrami operator.
We would like to stay in generality and look for solutions with $R$ being a constant. This is interesting as the consistency of the equations is kept provided the scalaron determines the off-diagonal elements of the metric. Thus, one has
\bea
&&\partial_\sigma\chi\partial^\sigma\chi+3M_{\rm p}^2 M^2\left(1-e^{-\sqrt{\frac{2}{3}}\frac{\chi}{M_{\rm p}}}\right)^2-M_p^2\mathcal{R}=0, \nonumber \\
&&-\frac{1}{\sqrt{-g}}\partial_\mu g^{\mu\nu}\sqrt{-g}\partial_\nu\chi=\sqrt{\frac{3}{2}} M_{\rm p} M^2\left(1-e^{-\sqrt{\frac{2}{3}}\frac{\chi}{M_{\rm p}}}\right)
e^{-\sqrt{\frac{2}{3}}\frac{\chi}{M_{\rm p}}}.
\eea
where $\Box_{\sigma}$ represents $\Box_{\sigma}=-\frac{1}{\sqrt{-g}}\partial_\mu g^{\mu\nu}\sqrt{-g}\partial_\nu$ and with the dimensionless rescaled field variable $\varphi=\sqrt{\frac{2}{3}}\frac{\chi}{M_{\rm p}}$, and $\varphi=-\log Z$, we obtain the following equation 
\bea
\label{eq:fullset2}
&&Z^{-2}\partial_\sigma Z\partial^\sigma Z+ 2M^2\left(1-Z\right)^2-\frac{2}{3}\mathcal{R}=0 \nonumber \\
&&-Z^{-1}\Box_g Z+Z^{-2}\partial_\sigma Z\partial^\sigma Z=M^2\left(1-Z\right)Z,
\eea
that inserted into the second obtaining yields finally
\be
\Box_g Z=-\left(2M^2-\frac{2}{3}\mathcal{R}\right)Z+3M^2Z^2-M^2 Z^3.
\ee
This is just a quartic scalar field theory as can be seen by changing the field variable to
\be
\label{eq:resc}
Z=1+\frac{\phi}{M},
\ee
that gives
\be
\label{eq:eom}
\Box_g\phi=-\left(M^2+\frac{2}{3}\mathcal{R}\right)M
+\left(M^2+\frac{2}{3}\mathcal{R}\right)\phi-\phi^3.
\ee

\medskip

\section{Quantization of the scalaron field}

From Ref.~\cite{Choudhury:2025iwg}, we know that the Dyson-Schwinger set of equations can be written down as
\bea
&&\partial^2 G_1(x)-\mu_R^2G_1(x)+[G_1(x)]^3+G_3(x,x,x)=\Omega, \nonumber \\
&&\\ \nonumber
&&\partial^2G_2(x,y)-\mu_R^2G_2(x,y)+3[G_1(x)]^2G_2(x,y)=\delta^4(x-y), \nonumber \\
&&\\ \nonumber
&&\partial^2G_3(x,y,z)-\mu_R^2G_3(x,y,z)+3G_1^2(x)G_3(x,y,z)+6G_1(x)G_2(x,y)G_2(x,z) \\ \nonumber
&&+3G_2(x,z)G_3(x,x,y)+3G_2(x,y)G_3(x,x,z) \\ \nonumber
&&+3G_2(x,x)G_3(x,y,z)+3G_1(x)G_4(x,x,y,z)+G_5(x,x,x,y,z)=0, \\ \nonumber
&&\\ \nonumber
&&\partial^2G_4(x,y,z,w)-\mu_R^2G_4(x,y,z,w)+3G_1^2(x)G_4(x,y,z,w)+6G_2(x,y)G_2(x,z)G_2(x,w)\\ \nonumber
&&+6G_1(x)G_2(x,y)G_3(x,z,w)+6G_1(x)G_2(x,z)G_3(x,y,w)+6G_1(x)G_2(x,w)G_3(x,y,z)\\ \nonumber
&&+3G_2(x,y)G_4(x,x,z,w)+3G_2(x,z)G_4(x,x,y,w)  \\ \nonumber
&&+3G_2(x,w)G_4(x,x,y,z)+3G_2(x,x)G_4(x,y,z,w) \\ \nonumber
&&+3G_1(x)G_5(x,x,y,z,w)+G_6(x,x,x,y,z,w)=0, \\ \nonumber
&\vdots&,
\eea
where now we have set
\be
\label{eq:muR1}
\mu_R^2=\mu_0^2-3G_2(x,x),
\ee
with the correction due to quantum fluctuations given by
\be
\mu_R^2=\mu_0^2-3\int\frac{d^4p}{(2\pi)^4}\frac{1}{p^2+2\mu_R^2},
\ee
that yields, after dimensional regularization, \cite{Choudhury:2025iwg}
\be
\label{eq:mu_R}
\mu_R^2=\frac{4 \pi ^2 \left(\sqrt{3} \sqrt{3 M^2+8 \mathcal{R}}+3 M\right)^2}{27 W\left(\frac{16}{27} \pi ^3 \left(\sqrt{3} \sqrt{3 M^2+8 \mathcal{R}}+3 M\right)^2 e^{-1+\gamma +\frac{16 \pi ^2}{3}}\right)},
\ee
where $\gamma$ is the Euler-Mascheroni constant.
From the properties of the quartic scalar field proven by Aizenman and coworkers \cite{Aizenman:1981zz,Aizenman:1982ze,Aizenman:2019yuo}, we can assume that the choice $G_k(x,x,\ldots) = 0$ for $k>2$, that is when two or more coordinates coincide, is a consistent one giving the expected Gaussian solution. Preserving translation invariance, we find the solution for $G_1$ from the equation
\be
-\mu_R^2G_1+G_1^3=0,
\ee
yielding $G_1=\pm\mu_R$, breaking spontaneously the conformal invariance of the theory by introducing a mass scale.

A recent study of the tunneling between different vacua was presented in Ref.~\cite{Calcagni:2022tls}.

In order to see the physical consequences of our result, it is not difficult to compute 
\be
G_2(p)=-\frac{1}{p^2-2\mu_R^2+i\epsilon},
\ee
where it is seen that the scalaron is massive and short ranged. This has some implications for the relevance of the ${\cal R}^2$ term in the low-energy limit and for its observability.

\medskip

\section{Contributions from matter fields}

We know from Ref.\cite{Shtanov:2022wpr,Pi:2017gih,Baez:2022stg,Samart:2018mow,Faraoni:2004pi,DeFelice:2010aj} that moving from the Jordan to the Einstein frame in the Starobisnky model or, more generally, for $f(\mathcal{R})$ theories implies a change in the energy-matter tensor in the equation of motion. One could ask how could possibly treated this situation. We can limit our analysis to the Higgs sector H written as 
\begin{equation} 
S_H^{(J)} = \int d^4x\,\sqrt{-g}\,\left[g^{\mu\nu}(D_\mu H)^\dagger (D_\nu H) - V_H(H)\right], 
\end{equation}
being
\begin{equation} 
V_H(H) = \lambda\left(H^\dagger H - \frac{v^2}{2}\right)^2. 
\end{equation}
where $\lambda$ is the self-quartic of the Higgs and $v$ is the vacuum expectation value (VEV). In the Starobinsky model one introduces the conformal transformation 
\begin{equation} 
\tilde{g}_{\mu\nu} = \Phi \, g_{\mu\nu}\, \qquad g_{\mu\nu} = \Phi^{-1}\tilde{g}_{\mu\nu}, 
\end{equation} 
changing the Higgs action to 
\begin{equation} 
S_H^{(E)} = \int d^4x\,\sqrt{-\tilde{g}}\left[
\Phi^{-1}\tilde{g}^{\mu\nu}(D_\mu H)^\dagger (D_\nu H)
-\Phi^{-2}V_H(H) \right]. 
\end{equation}
In our case we have
\begin{equation} 
\Phi = e^{\sqrt{\frac{2}{3}}\frac{\chi}{M_{\rm Pl}}}, 
\end{equation} 
giving in the Einstein frame for the kinetic term
\begin{equation} 
\mathcal{L}_H = e^{-\sqrt{\frac{2}{3}}\frac{\chi}{M{\rm Pl}}}\tilde{g}^{\mu\nu}(D_\mu H)^\dagger (D_\nu H)
e^{-2\sqrt{\frac{2}{3}}\frac{\chi}{M_{\rm Pl}}}V_H(H), 
\end{equation}
and for the potential
\begin{equation} 
V_H(H) = \lambda\left(e^{-\sqrt{\frac{2}{3}}\frac{\chi}{M_{\rm Pl}}}\hat{H}^\dagger\hat{H} - \frac{v^2}{2}\right)^2, 
\end{equation} 
while the conformal prefactor yields 
\begin{equation} 
e^{-2\sqrt{\frac{2}{3}}\frac{\chi}{M_{\rm Pl}}}V_H(H) = \lambda\left(\hat{H}^\dagger\hat{H} 
- \frac{v^2}{2}e^{-\sqrt{\frac{2}{3}}\frac{\chi}{M_{\rm Pl}}}\right)^2. 
\end{equation}
The breaking of the conformal invariance seen in the preceding section,
granting the solution $\chi=\text{constant}$ arising from the 1P-function $G_1$,
will keep the shape of the Higgs sector invariant as the conformal factor becomes just a constant at the transition.

\medskip

\section{Non-minimal coupling case}

We consider the following Lagrangian for the non-minimal coupling case
\be
\label{eq:lagrangian}
\mathcal{L}= 
\frac{1}{2}(\partial \phi )^2-V(\phi)+\frac{1}{2}\frac{\mathcal{R}}{\kappa}\left(1-\xi \frac{12}{\Lambda} \phi^2 \right),
\ee
where $\phi$ is the scalar field, $\kappa=8\pi G$ being $G$ the Newton constant, $\mathcal{R}$ is the Ricci scalar, $\Lambda$ is the cosmological constant and $\xi$ is the non-minimal coupling and the potential is taken to be in the simplest conformal invariant form
\be
V(\phi)=\frac{\lambda}{4}\phi^4.
\ee
Starting from the Lagrangian (\ref{eq:lagrangian}), we can write down the set of Dyson-Schwinger equations \cite{Frasca:2015yva}
\bea
&&\partial^2 G_1(x)+\lambda\left([G_1(x)]^3+3G_2(x,x)G_1(x)+G_3(x,x,x)\right)+\xi \mathcal{R} G_1(x)=0 \nonumber \\
&& \nonumber \\
&&\partial^2G_2(x,y)+\lambda\left(3[G_1(x)]^2G_2(x,y)+3G_2(x,x)G_2(x,y)\right. \nonumber \\
&&\left.+3G_3(x,x,y)G_1(x)+G_4(x,x,x,y)\right)+\xi \mathcal{R} G_2(x,y)=\delta^4(x-y) \nonumber \\
&&\nonumber \\
&&\partial^2G_3(x,y,z)+\lambda\left[6G_1(x)G_2(x,y)G_2(x,z)+3G_1^2(x)G_3(x,y,z)\right. \nonumber \\
&&\left.+3G_2(x,z)G_3(x,x,y)+3G_2(x,y)G_3(x,x,z)\right.  \nonumber \\
&&\left.+3G_2(x,x)G_3(x,y,z)+3G_1(x)G_4(x,x,y,z)+G_5(x,x,x,y,z)\right]+\xi \mathcal{R} G_3(x,y,z)=0 \nonumber \\ 
&& \nonumber \\
&&\partial^2G_4(x,y,z,w)+\lambda\left[6G_2(x,y)G_2(x,z)G_2(x,w)\right. \nonumber \\
&&+6G_1(x)G_2(x,y)G_3(x,z,w)+6G_1(x)G_2(x,z)G_3(x,y,w)  \nonumber \\ 
&&+6G_1(x)G_2(x,w)G_3(x,y,z)+3G_1^2(x)G_4(x,y,z,w) \nonumber \\
&&+3G_2(x,y)G_4(x,x,z,w)+3G_2(x,z)G_4(x,x,y,w) \nonumber \\ 
&&+3G_2(x,w)G_4(x,x,y,z)+3G_2(x,x)G_4(x,y,z,w) \nonumber \\
&&\left.+3G_1(x)G_5(x,x,y,z,w)+G_6(x,x,x,y,z,w)\right]+\xi \mathcal{R} G_4(x,y,z,w)=0 \nonumber \\ 
&\vdots&,
\eea
where $\partial^2=(1/\sqrt{-g})\partial_\alpha\sqrt{-g}g^{\alpha\beta}\partial_\beta$ is the Laplace-Beltrami operator and $R=\frac{12\mathcal{R}}{\kappa\Lambda}$. We want to evaluate the 1P- and 2P-correlation functions for some choice of the metric. Working in the same limit as for the Starobinsky model with the Ricci scalar constant and very large, we will get in this case the equation for $G_1$
\be
\label{eq:xiR}
\lambda G_1^3+3\lambda G_2(x,x)G_1+\xi \mathcal{R}G_1=0.
\ee
We observe that, after regularization, $G_2(x,x)$ takes on negative values (see eq.(\ref{eq:mu_R})
but, looking at the solutions for the minima of the potential in eq.(\ref{eq:xiR}),
the presence of the non-minimal term can impede the phase transition to occur. This is consistent with Ref. \cite{Calcagni:2022tls} where it was realized that the general effect of the presence of such a non-minimal coupling $\xi$ is to hinder the tunneling between the vacua.

\medskip

\section{Discussion and Conclusion\label{sec5}}
In this review we recapped our methods to tackle the strongly coupled theories, particularly involving gravity with backgrounds like de Sitter, quadratic gravity and non-minimally coupled scalar to Ricci curvature $\mathcal{R}$. Under certain conditions, very similar to what happens in Nambu-Jona-Lasinio model which can be used to describe strong interactions \cite{Klevansky:1992qe}, an effective description involving a non-linear $\sigma-$ renormalizable model can be achieved, here too we were able to show a scalar to describe the gravity. Particularly we show that starting from a simple model of quadratic gravity involving the Starobinsky term $\mathcal{R}^2 + \mathcal{R}$, the scenario in the strong coupling limits boils down to having a large and constant $\mathcal{R}$ along with mass gap generated for the scalaron field. The strong coupling condition can be understood as $3M^2<\mathcal{R}$, being $M$ the mass term due to the quadratic term leading to self interaction among the scalaron. This reflects the fact that this term represents the self-interactions among the gravitons and the scalar degrees of freedom in the theory. Particularly, we found that in a theory involving no cosmological constant for $\Lambda=0$, the theory possesses a mass gap. Both for the Starobinsky and non-minimally coupled scenarios, we get an effective potential description which involves a false vacuum like behaviour and may undergo phase transitions. \emph{Particularly,} in the scenario involving the non-minimal coupling to gravity, the strong coupling as we know impacts primordial cosmological scenarios for instance those involving strong first-order phase transitions, leading to the search of a stochastic gravitational-wave background from in early universe. The LIGO--Virgo--KAGRA network already looks for such a signal \cite{Romero:2021kby,LIGO-result}, and leaves open the window to develop and test techniques for vacuum decay with GW signatures in the strongly coupled limits. We envisage that this new technique to explore the non-perturbative window could be tested in near future with current and future planned worldwide network of GW detectors.

\medskip

A.G. acknowledges the support from the Royal Society Fellowship, funding Reference: NIF\textbackslash R1\textbackslash 253963.

\end{document}